\documentstyle[preprint,12pt,aps,epsf]{revtex}

\tighten
\begin{document}
\draft
\preprint{Imperial/TP/97-98/5}

\newcommand{\nc}{\newcommand}
\nc{\al}{\alpha}
\nc{\ga}{\gamma}
\nc{\de}{\delta}
\nc{\ep}{\epsilon}
\nc{\ze}{\zeta}
\nc{\et}{\eta}
\renewcommand{\th}{\theta}
\nc{\ka}{\kappa}
\nc{\la}{\lambda}
\nc{\rh}{\rho}
\nc{\si}{\sigma}
\nc{\ta}{\tau}
\nc{\up}{\upsilon}
\nc{\ph}{\phi}
\nc{\ch}{\chi}
\nc{\ps}{\psi}
\nc{\om}{\omega}
\nc{\Ga}{\Gamma}
\nc{\De}{\Delta}
\nc{\La}{\Lambda}
\nc{\Si}{\Sigma}
\nc{\Up}{\Upsilon}
\nc{\Ph}{\Phi}
\nc{\Ps}{\Psi}
\nc{\Om}{\Omega}
\nc{\ptl}{\partial}
\nc{\del}{\nabla}
\nc{\be}{\begin{eqnarray}}
\nc{\ee}{\end{eqnarray}}
\nc{\bh}{{\bf H}}
\nc{\cm}{{\cal M}}
\nc{\gu}{\Ga_U(2)}

\title{On The Beta--Function in N=2 Supersymmetric Yang-Mills Theory}
\author{Adam Ritz\footnote{email: a.ritz@ic.ac.uk}\\ $\;$\\}
\address{Theoretical Physics Group, Blackett Laboratory,  \\
         Imperial College, Prince Consort Rd.,
         London, SW7 2BZ, U.K.}
\date{\today}

\maketitle

\begin{abstract} 
The constraints of $N=2$ supersymmetry, in combination with several other
quite general assumptions, have recently been used to show that $N=2$ 
supersymmetric Yang-Mills theory has a low energy 
quantum parameter space symmetry characterised by the discrete group $\gu$.
We show that if one also assumes the commutativity of renormalization group
flow with the action of this group on the complexified coupling
constant $\ta$, then this is sufficient to determine the non-perturbative 
$\beta$-function, given knowledge
of its weak coupling behaviour. The result coincides with
the outcome of direct calculations from the Seiberg-Witten solution.
\end{abstract}


\section{Introduction}

In recent years there has been remarkable progress in understanding
the non-perturbative dynamics of supersymmetric gauge theories.
The powerful holomorphy constraints imposed by supersymmetry (SUSY)
have allowed results, calculable at weak coupling, to be analytically
continued to the strong coupling regime along quantum-mechanically
flat directions in the space of inequivalent vacua (see e.g. \cite{is}). 
In particular,
the constraints of $N=2$ SUSY have allowed an exact solution for
the holomorphic two-derivative contribution to the Wilsonian
effective action to be determined by Seiberg and Witten \cite{sw94}
for $N=2$ supersymmetric Yang-Mills (SYM) theory and SQCD. More recently,
it has been argued \cite{bmt,fmrss} that
a number of the conjectures made in \cite{sw94}, in particular 
those of physical importance such as
electric-magnetic duality, are strictly unnecessary in order to
obtain a unique solution. Indeed the 
existence of an underlying discrete parameter-space symmetry group of
the full quantum theory, by which we mean transformations the couplings 
of the theory which leave the vacuum state and full mass spectrum
invariant, has been determined uniquely from several rather
general requirements including unbroken $N=2$ supersymmetry \cite{fmrss}.

Quite generally, the existence of such parameter space symmetries is
of great utility, as they act on the same space as renormalization
group (RG) transformations of the theory, and thus place restrictions
on the structure of RG flow. In particular, when discrete symmetries
associated with a subgroup of the modular group $SL(2,${\bf Z}$)$
hold at all scales in a quantum theory, then the $\beta$--function
must satisfy certain modular transformation properties. When the symmetry
group is large enough, for example $SL(2,${\bf Z}$)$ itself, the constraint
on the $\beta$--function is generally strong enough to force it to vanish,
rendering the theory scale invariant. However, in cases where the symmetry
group is smaller, the RG flow may still be nontrivial albeit highly
constrained. In such cases, study of the required modular transformation
properties of $\beta$ often provides nonperturbative information about
the RG flow. Such arguments have been used to highly constrain the 
structure of the RG $\beta$--functions in statistical systems \cite{statm}, 
and also the non-linear sigma model \cite{nlsm}.
 
In this letter, we re-analyse renormalization group flow in $N=2$ SYM
from this perspective of parameter space symmetries. A particular
nonperturbative definition of the $\beta$--function, associated with
the flow of the couplings along the moduli space, arises naturally
in this context. This $\beta$--function has been obtained
previously from the Seiberg--Witten solution by Minahan and 
Nemeschansky \cite{mn}, and also Bonelli and Matone \cite{bm}.
Here we show that this result follows from the following conditions,
along with knowledge of its weak-coupling behaviour. These conditions,
although not in their most basic form, are given by:
\begin{enumerate}
\item The maximal parameter-space symmetry group 
      of the low energy effective theory,
      when acting on the complexified coupling $\ta$,
      is given by $\gu$\footnote{Throughout we shall use the notation
      of Rankin \cite{rankin} for subgroups of $SL(2,${\bf Z}$)$ and their
      generators. Note that $\gu$ is also commonly
      denoted $\Ga_0(2)$.}, the index--3 subgroup of 
      $PSL(2,${\bf Z}$)=SL(2,${\bf Z}$)/${\bf Z}$_2$.
\item This equivalence group commutes, in an appropriate sense to be defined
      below, with the flow of $\ta$ induced by the renormalization group.
\end{enumerate}
In order to justify condition 1, we also require the conditions
shown in \cite{fmrss} to be sufficient to ensure the uniqueness of this
symmetry group. We shall discuss these additional assumptions shortly,
although we note that no assumption about the existence
of a dual magnetic description at strong coupling will be required. 

The second condition above is crucial, and is the statement of compatibility
between the action of the symmetry group and the renormalization group flow.
Once satisfied, it implies that if the parameter space symmetry
holds at one scale, then it will be preserved by the RG flow, and continue
to hold at lower scales. 
 
In Section 2, after briefly reviewing some general features
of the effective theory, we recall the conditions
imposed in \cite{fmrss} which were used to ensure condition 1. 
Although it is not clear at this stage if all these assumptions
are strictly necessary, certain holomorphy conditions on the coupling, 
following from unbroken $N=2$ SUSY, and the allowed 
singularity structure structure over the moduli space of vacua, 
will be important in the following analysis. In Section 3 we discuss the
precise characterization of the renormalization group flow to be studied, 
and give a definition of $\beta(\ta)$. We then illustrate how condition 2
allows determination of $\beta(\ta)$ where all unknown parameters 
may be fixed purely via knowledge of the behaviour in the
semi-classical limit. 

\section{The maximal equivalence group}

Unbroken supersymmetry ensures that the potential for the
scalar component $\ph$ of the $N=2$ multiplet vanishes, even when 
quantum corrections are included, when $\ph$ takes values in the
Cartan subalgebra of the gauge group, which we shall take to be $SU(2)$.
Classically, for non-zero $a\equiv\left<\ph\right>$ 
the Higgs mechanism spontaneously
breaks the gauge symmetry to $U(1)$, with the arbitrariness in 
$\left<\ph\right>$,
or more precisely a gauge invariant parameter such as 
$\left<{\rm Tr}\ph^2\right>$,
leading to a moduli space of vacua $\cm$.
It is a consequence of $N=2$ supersymmetry that the structure
of the low energy Wilsonian effective action for the light $U(1)$
multiplet may be represented in terms of an $N=2$ superfield 
${\cal A}=\ph+\th\ps+\cdots$ as \cite{seiberg88}
\be
 \Ga_W[{\cal A}] & = & \frac{1}{4\pi} {\rm Im} \int
     {\rm d}^4x{\rm d}^2\th_1{\rm d}^2\th_2{\cal F}({\cal A})+\cdots,
\ee  
up to non-holomorphic higher derivative terms, where
${\cal F}$ is the holomorphic prepotential. In $N=1$ superspace
this action takes the form
\be
 \Ga_W[A,W_{\al}] & = & \frac{1}{4\pi}{\rm Im}\int {\rm d}^4x
 \left[\int{\rm d}^4\th K(A,\overline{A})+\int{\rm d}^2\th\frac{1}{2}
      \ta(A)W^{\al}W_{\al}\right]+\cdots,
\ee
where $A$ and $W_{\al}$ are $N=1$ chiral and vector superfields,
$K\equiv{\cal F'}(A)\overline{A}$ is the K\"ahler potential, and
\be
	\label{taudef}
 \ta(A) & \equiv & \frac{\ptl^2 {\cal F}}{\ptl A^2}.
\ee
Since $\ta(A)$ is the coefficient of the kinetic term, its imaginary
part must be positive. Its real part also plays a role similar to the
theta parameter in the microscopic theory, and thus it is natural to
define the corresponding effective parameters in the manner
\be
 \ta(u) & \equiv & \frac{2\pi}{\theta_{eff}(u)}+\frac{4\pi i}{g^2_{eff}(u)},
      \label{tau}
\ee
where $u$ is a gauge invariant parameter labelling the moduli
space. Careful analysis in \cite{fp} has shown that one may match these
effective parameters to the underlying microscopic $SU(2)$ parameters
at small scales via an appropriate identification of the $\La$ scale of the
effective theory with the renormalization-scheme-dependent
dynamically generated microscopic scale.

Although we shall need the relationship only in the weak--coupling region, 
it has been shown quite generally \cite{uid}
that for the pure Yang-Mills case\footnote{The situation is more complex
for certain models with higher matter field content \cite{dkm2}.} 
$u$ may be identified
as $u=\left<{\rm Tr}\ph^2\right>$. Thus  $\left<{\rm Tr}\ph^2\right>$
is a good global coordinate on the moduli space, as was conjectured in 
\cite{sw94}. The point is that 
$a$, when defined as the expectation value of the scalar component of
the $N=2$ superfield ${\cal A}$, only corresponds to a useful
parametrisation in the semi-classical domain.

An important insight is that the moduli space coordinatized
by $u$, may also be parametrized in terms of
$\ta$ which plays the role of a convenient uniformizing parameter.
This requires knowledge of the multivaluedness of the relation $\ta=\ta(u)$. 
While, for unitarity $\ta \in\;$ {\bf H}, 
the upper-half complex plane, it is clear from the presence of
an effective theta parameter in ({\ref{tau}) that there exist equivalence
relations between various values of $\ta$. In other words the mapping
between $\ta$ and $u$ is not one-to-one. In particular, one
may identify $\ta \mapsto U(\ta)=\ta+1$, corresponding to a rotation of
the effective $\theta$-angle by $2\pi$. Following
\cite{fmrss}, we define the {\it maximal equivalence group} $G$ as the
group of all transformations of $\ta$ which leave the vacuum, $u(\ta)\in \cm$,
and the full mass spectrum, invariant. Since $g^2_{eff}$ must be 
positive, this ensures that $G\subset SL(2,${\bf R}$)$. In order to
constrain $G$ further, we note that $N=2$
SUSY implies that the mass of charged particles is BPS 
saturated \cite{wo}, i.e. $M=\sqrt{2}|Z|$, where $Z$ is the SUSY central 
charge. It has been argued in \cite{sw94,cru,mors} that quantum mechanically
this relation has the form
\be
 M & = & \sqrt{2}|q_e a+q_m a_D|,
\ee
where $a_D\equiv{\cal F}'(a)$, and $q_e$
and $q_m$ are integer valued charges.
The invariance of this spectrum, and the fact that $\ta$ satisfies 
$\ta=da_D/da$ as follows from (\ref{taudef}), then ensures that $G$ is a 
subgroup of $PSL(2,${\bf Z}$)$, and thus
the physical moduli space may be represented as the fundamental
$\ta$--domain\footnote{Note that a more general definition
is necessary when one adds matter hypermultiplets \cite{nahm,mors}.} 
$D=${\bf H}$/G$.  

Having presented evidence above for the nontriviality of 
$G\subset PSL(2,${\bf Z}$)$,
we now recall the analysis of \cite{fmrss} which argued that 
the following assumptions constitute a set of 
sufficient conditions to determine $G$ uniquely:
\begin{enumerate}
\item $\ta$ takes all values in {\bf H}.
\item There are a finite number of singular points in $\cm$.
\item The BPS mass $M$ is single-valued on $\cm$.
\item The mass of the lightest charged field is finite except
       in the perturbative region. 
\end{enumerate}
With these assumptions, it was shown that $G$ is given 
by $\gu\subset\Ga(1)=SL(2,${\bf Z}$)$, and thus these requirements
also serve as sufficient conditions to ensure the validity of our initial
assumption that the symmetry group of the quantum theory is $\gu$.
Conditions 2 and 4 deserve further comment. In particular, condition
2 ensures that functions, or more generally sections, defined over 
the moduli space, will be meromorphic. Since singular points in the moduli
space will be mapped to the vertices of the fundamental domain of
$G$, this ensures that there are no singularities in the interior
of this fundamental domain. Meanwhile,
condition 4 is the expected behaviour for a theory with only one
asymptotically free regime, as we observe in the present case.

\section{Determination of the Beta--Function}

The above definition of the equivalence group, as acting at a fixed point of 
the moduli space, is important in determining the precise characterisation
of RG flow we shall use. In particular,
in order to consistently apply the second assumption of Section 1, 
we need to consider the infinitesimal
flow at a {\it fixed} point of the moduli space. Since the effective theory
has been constructed via recourse only to the general constraints
imposed by supersymmetry, there is no explicit cutoff scale, and 
the only dimensionful parameters of the theory are $u$ (or $a$) and $\La$. 
In order to allow compatibility with the action of the equivalence group 
it is then appropriate, following \cite{mn,rgeqn,bm}, to define 
the $\beta$--function for the evolution of $\ta$ as
\be
 \beta(\ta) & \equiv & \La \left.\frac{\ptl \ta}{\ptl \La}\right|_{u}
      = - 2\left.\frac{u}{u'}\right|_{\La},
     \label{betdefinit}
\ee
where $u'$ denotes $\ptl u/\ptl \ta$. The latter relation,
noted in \cite{bm}, follows
from the fact that, since $\ta$ is dimensionless, 
$\beta=\beta(u/\La^2)$ and the above definition is physically 
equivalent to considering the flow induced by motion over the 
moduli space, i.e. changing $u$ with $\La$ fixed. 
The latter description is perhaps more physically relevant, and closer
in spirit to a more standard weak-coupling definition such
as \cite{seiberg88}, $\beta^{a} \equiv a\ptl_a \ta|_{\La}$, where the
renormalization scale is chosen equal to the vev.
The choice of $u$ as the scale in the nonperturbative
definition (\ref{betdefinit}) is motivated by the fact that $u$ is
a more appropriate coordinate for the moduli space at strong coupling.
However, while these definitions may differ in the strong coupling
region one expects that they should be equivalent, at least up to an overall
constant, in the perturbative region $a\gg \La$. This may be
verified by considering the appropriate differentials of $a$, $u$, and 
$\ta$ \cite{bm}, from which one obtains the following relation,
\be
 \beta & = & \beta^{a}\left(\La\left.
      \frac{\ptl}{\ptl \La}\ln a\right|_u-1\right)
   \;\rightarrow\; -\beta^{a}+\cdots.
\ee
The final limit is taken in the perturbative region, where the relation 
simplifies as expected since in this case $a\sim\sqrt{2u}+\cdots$ which
is independent of $\La$. Thus perturbatively the definitions agree
up to a sign, while non-perturbatively there is a discrepancy due 
to the fact that $a$ is not a good global coordinate on the moduli space.

At weak coupling, we may match this $\beta$--function for
the effective coupling to the running of the microscopic coupling.
The 1--loop contribution is then given by\footnote{Note that there 
is a relative minus sign compared to \cite{hsw,seiberg88} 
due to the definition of the $\beta$--function
in (\ref{betdefinit}).} $\beta_{pert}(\ta)=-2i/\pi$ \cite{hsw}.
Note that there are no higher loop contributions
\cite{gsr,nsvz,seiberg88}, but instantons \cite{nsvz} lead to additional 
nonperturbative effects as first discussed for N=2 theories by Seiberg 
in \cite{seiberg88}. It is these contributions to $\beta(\ta)$ that we
shall determine below.

It is important to note that $\beta(\ta)$ inherits certain
analyticity properties from those of $\ta(u)$. The holomorphy of
$\ta$ follows from its definition as the derivative of ${\cal F}$ and
the existence of unbroken supersymmetry. Furthermore, as was
noted in the previous section, since possible singular points
in the moduli space are mapped to the vertices of the fundamental 
domain $D$, $\beta(\ta)$ must be regular in the interior of this domain.  

We now concentrate on the global definition of $\beta$ given in Eq. 
(\ref{betdefinit}) and make use of the second assumption of Section~1
in order to obtain an explicit expression. Recall that
the existence of a nontrivial equivalence group $\gu$ implies that
the physical moduli space $\cm$ reduces to the corresponding fundamental
domain $D=$\bh$/\gu$. The action of a general element of this 
equivalence group, $\ga \cdot \ta=(a\ta+b)/(c\ta+d)$, where $ad-bc=1$, 
$a,b,c,d\in\;${\bf Z}, may be conveniently
represented in terms of the generators $U$ and $VU^2V$ \cite{rankin},
\be
 U: \ta &\rightarrow& \ta+1\;\;\;\;\;\;\;\;\;
     VU^2V: \ta \rightarrow \frac{\ta}{1-2\ta}.
\ee
The assumed validity of this symmetry at all scales leads to
the condition that the RG flow commute with
this action of the equivalence group. From the restrictions on its
allowed analytic structure, this implies that $\beta(\ta)$
for $\ta\in\;$\bh, satisfies \cite{statm}
\be
 \beta(\ga \cdot \ta) &=& \frac{d(\ga\cdot\ta)}{d \ta}\beta(\ta) = 
       (c\ta+d)^{-2}\beta(\ta),
\ee
where $\ga\in\gu$. That is to say, $\beta(\ta)$ transforms as
a modular form of $\gu$ of weight $-2$. Note that the fact that $\beta$ is not
invariant under such transformations is due to its definition
as a contravariant vector field on the space of couplings.

Positive weight modular forms may be obtained constructively
in terms of generalised Eisenstein series. However, for negative weight
forms no such construction exists. Nevertheless, we may use
a general theorem, valid for all even weight forms \cite{rankin},
which states that any weight $-2$ modular form of a discrete group
$\Ga \subset \Ga(1)$, 
may be represented in terms of a univalent
automorphic function $f$ of $\Ga$, via
\be
 \beta(\ta) & = & \frac{1}{f'}\frac{P(f)}{Q(f)}, \label{betdef}
\ee
where $f'$ denotes $\ptl f/\ptl \ta$, and $P$ and $Q$ are polynomials
in $f$. In the present case $\Ga=\gu$, and a convenient automorphic
function is given by $f=F(\ta)=-\th_3^4\th_4^4/\th_2^8$, where 
$\th_i=\th_i(0|\ta)$, for $i=2,3,4$, are the theta constants.

We could now proceed to place constraints on the polynomials $P$ and $Q$
from the known weak-coupling asymptotics. However, an alternative approach is to
recall that the definition of the $\beta$--function
(\ref{betdefinit}), when expressed in terms of $\tilde u=u/\La^2$, has 
precisely the form (\ref{betdef}), with $\tilde u$ playing the role of $f$. 
Furthermore, since
$\tilde u$ is clearly automorphic from the definition of the equivalence group,
and univalent due to its parametrisation of the moduli space, we may
equally well identify $f=\tilde u(\ta)$.

Obtaining an explicit expression for (\ref{betdef}) then reduces
to determining the relation $\tilde u=\tilde u(\ta)$ for the automorphic function $\tilde u$.
Using only the required transformation properties under $\gu$, and
the known weak coupling perturbative asymptotics for $\ta(u\sim a^2/2)$
arising from the 1--loop $\beta$--function, this relationship was
obtained by Nahm \cite{nahm}. These are the assumptions of the present
paper, and thus we may use this result which, in corrected form and
with a convenient normalization of the scale $\La$, reads \cite{nahm}:
\be
 \left(\frac{u}{\La^2}\right)^2 & = & 1-4F. \label{uf}
\ee

This simple relationship between $F$ and $\tilde u^2$ is essentially 
demanded by their required transformation properties under $\gu$.  To gain
a little more insight into this expression, we may also recover
the functional relation, $F(\tilde u)$, as follows. 
Since $\tilde u$ and $F$ are both univalent and automorphic under
$\gu$, they may be functionally related by a polynomial of degree
determined by their singularity structure \cite{rankin}. 
Importantly, since the singularity of $\tilde u$ is at the weak--coupling  
vertex, $\ta=i\infty$, of $D$, we may extract the 
order of the pole from weak coupling asymptotics. Introducing
an elliptic modulus $k^2(\ta)=\th_2^4/\th_3^4$ which has a zero
at $\ta=i\infty$, and using the weak coupling
relation for $\ta(u\sim a^2/2)$, obtained by integrating the 1--loop
$\beta$--function, we find that $u\rightarrow 2k^{-2}$
at the weak coupling vertex. Similarly, using the explicit
representation for $F$ in terms of complete elliptic integrals
\cite{gr}, we find $F \rightarrow -k^{-4}$. Thus $F$
has a pole of order 2 at the pole of $\tilde u$, and consequently we may 
write \cite{rankin}
\be
 F & = & c_1\left(\frac{u}{\La^2}\right)^2 
    + c_2\left(\frac{u}{\La^2}\right)+ c_3. \label{reln}
\ee
The univalence of $F$ and $\tilde u$ implies that $c_2=0$, while
the perturbative asymptotics, arising from the 1-loop 
$\beta$--function, $\beta_{pert}(\ta)=-2i/\pi$ \cite{hsw}, 
implies $c_1=-1/4$. From the earlier discussion, we now expect that
instanton contributions to $\beta$ are determined by the value of $c_3$.

Since $F$ and $\tilde u$ are univalent, we may fix $c_3$ by a choice
of the zero. In the analysis of \cite{nahm} discussed above, it was
pointed out that the zero of $\tilde u(\ta)$ should lie at the orbifold vertex
of $\gu$, $\ta=(i-1)/2$. This fixes $c_3=1/4$ and (\ref{reln}) then reduces
to (\ref{uf}). 

More generally, the zero of $\tilde u(\ta)$ may not be fixed by the group 
structure and thus, from a calculational point of view, it is helpful
to consider an expansion of this result near the weak coupling vertex, 
without first fixing $c_3$,
\be
 \beta & = & \frac{4c_3-4F}{F'} \sim 
     -\frac{2i}{\pi}\left(1+\frac{1}{32}
     \left(32c_3-3\right)k^4+\cdots\right). \label{expan}
\ee
The $O(k^4)$ correction may be associated with
a 1-instanton contribution \cite{seiberg88}. This may be seen by noting that
$k^2\sim 2\exp(i\pi\ta)+\cdots$ in this limit and therefore,
setting the $\theta$-angle to zero for clarity, we have
\be
 \beta & \sim & -\frac{2i}{\pi}\left(1+\frac{1}{8}
     \left(32c_3-3\right)\exp\left(\frac{-8\pi^2}{g^2(u)}\right)+\cdots\right),
\ee 
which exhibits the standard 1-instanton exponential factor.

As a consequence, we may also fix the constant $c_3$ from knowledge of
the 1-instanton correction which is calculable at weak coupling.
In effect, when restricted to the 1-instanton level,
the constant $c_3$ has implicitly been fixed by the choice of perturbative
renormalization scheme. In order to see
this more clearly we note that in the weak coupling limit we can identify $\La$
with the perturbative renormalization group invariant scale, given by
$\La^4=u^2\exp(2i\pi\ta(u))$. Thus we have  
$k^4\sim 4\La^4/u^2 +\cdots$
and, from the structure of (\ref{expan}), we observe
that a change of renormalization scheme corresponding to a change in 
$\La$ may be compensated by a 
change in $c_3$ at the 1--instanton level. The 1--instanton term is therefore
scheme dependent and a choice of scheme fixes the coefficient of
the associated exponential factor, and thus the value of $c_3$, unambiguously.

Therefore, the final constant may be
fixed via knowledge of the 1-instanton contribution
at weak coupling, calculable by saddle point methods in a scheme
such as Pauli-Villars. Such a scheme was shown in \cite{fp} to 
be equivalent to the implicit scheme used in \cite{sw94}.
However, a direct instanton calculation \cite{fp} gives the first
power correction to the perturbative result at a fixed value of $a$, rather
than $u$. 
Nevertheless, while $\beta|_u$ and $\beta|_a$ differ at this order,
the relationship, $u=a^2/2+\La^4/(4a^2)$, is again calculable
at weak coupling to the required 1-instanton order using the fact
that\footnote{Our normalisation
of $\La$ differs from that of \cite{fp} by a factor of $\sqrt{2}$. i.e.
$\La^2=2\La_{\overline{DR}}^2$, which fixes the renormalization scheme.} 
$u=\left<{\rm Tr}\ph^2\right>$. Thus, converting from $u$ to $k$, the 
1--instanton induced power correction to $\beta$ takes the form
\be
 \beta_{\mbox{weak coupling}} & = & 
      -\frac{2i}{\pi}\left(1+\frac{5}{32}k^4+\cdots\right), \label{1inst}
\ee
where the dots represent higher order instanton contributions.
Comparing (\ref{expan}) with (\ref{1inst}) leads to the
identification $c_3=1/4$, consistent with our earlier conclusion.

The final result for the $\beta$--function is then given by
\be
 \beta(\ta) &=& \frac{1-4F}{F'} = -\frac{i}{\pi} 
     \left(\frac{1}{\th_3^4}
         +\frac{1}{\th_4^4}\right). \label{bet}
\ee
After accounting for the alternative $\theta$--function
notation used, one may readily verify that this result coincides
with that obtained by Minahan and Nemeschansky \cite{mn} 
from the elliptic curve of the
Seiberg--Witten solution. This expression may also be
shown to coincide with the result obtained by Bonelli and Matone \cite{bm}
from the Picard--Fuchs equation for the vevs $a$ and $a_D$
of the chiral superfield. However, this requires use of
an alternative choice of boundary conditions in this equation
in order to be consistent with the choice $k^2=\th_2^4/\th_3^4$. 
The singularity structure of this $\beta$--function has 
been discussed previously in \cite{mn} and \cite{bm}. We recall here that
the fixed points, located at $\ta=(i-1)/2$, and the equivalent points 
under $\gu$, correspond to $u=0$, where
the full gauge symmetry is classically restored. This result is
not unexpected recalling that $\beta=\beta(u/\La^2)$. Finally we note
that the $\beta$--function is
singular at the vertices of the fundamental domain which lie on the
real axis, i.e. $\ta\in\;${\bf Z} in general. The gauge coupling diverges
at these points, which are associated with a breakdown of the effective
theory due to the presence of extra massless monopoles and dyons.

Finally, we note that one may expand (\ref{bet}) to higher order allowing
comparison of the 2-instanton coefficient with semi-classical calculations. 
Expanding (\ref{bet}) up to
$O(k^8)$, which includes all 2--instanton effects plus partial
corrections due to three instantons, we obtain
\be
 \beta & \sim & -\frac{2i}{\pi}\left(1+\frac{5}{32}k^4+\frac{5}{32}k^6
     +\frac{1229}{8192}k^8+\cdots\right).
\ee
In order to compare this result with saddle-point calculations, which
are obtained at fixed $a$, rather than $u$, it is necessary to 
carefully convert the 2-instanton results \cite{dkm} for ${\cal F}$
(or $\ta$) to functions of $u$ or $k$ (see e.g. \cite{klt}).  
Once one does this, and accounts for the different normalisation 
of $\La$ in \cite{dkm}, one readily verifies the explicit
2-instanton coefficient obtained by evaluation of the induced vertex.

\subsection*{Acknowledgements}
I would like to thank D.~Z. Freedman, J.~I. Latorre, D.~F. Litim, 
A.~A. Tseytlin, and A.~I. Vainshtein, for helpful comments and correspondence,
and the financial support of the Commonwealth Scholarship Commission
and the British Council is also gratefully acknowledged.

\bibliographystyle{prsty}

\end{document}